\documentclass{moriond}

\def\be{\begin{equation}}
\def\ee{\end{equation}}
\def\bea{\begin{eqnarray}}
\def\eea{\end{eqnarray}}

\newcommand{\Photo}{\includegraphics[height=35mm]{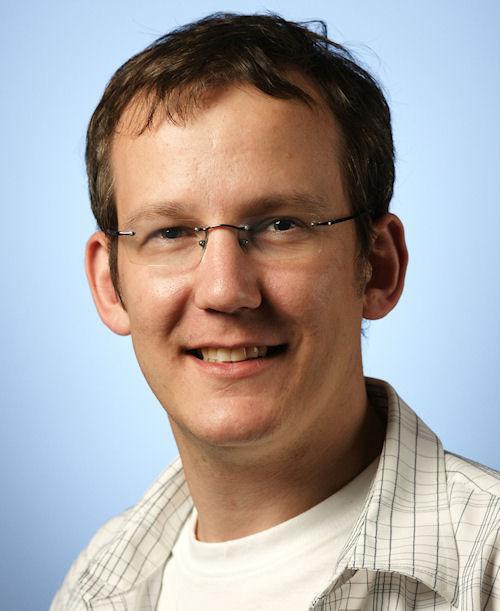}}
\begin{document}
\vspace*{0cm}
\title{BINGO - A novel method to detect BAOs using a total-power radio telescope}

\author{Clive Dickinson\,\footnote{On behalf of the BINGO collaboration.}}

\address{Jodrell Bank Centre for Astrophysics, School of Physics \& Astronomy, Alan Turing Building, \\
The University of Manchester, Oxford Road, Manchester, M13 9PL, U.K.}

%%%%%%%%%%%%%%%%%%%%%%%%%%%%%%%%%%%%%%%%%%%%%%%%%%

\maketitle\abstracts{
BINGO is a novel single-dish total-power telescope that will map the redshifted HI sky in a $\sim 15^{\circ}$ strip, at frequencies of 960--1260\,MHz ($z=0.12$--0.48). BINGO will have the sensitivity to accurately measure the HI power spectrum and to detect Baryon Acoustic Oscillations (BAOs) for the first time at radio wavelengths. This will provide complementary cosmological information to existing surveys and will measure the acoustic scale to $\approx 2\,\%$ precision. We provide an update on BINGO including an improved two-mirror optical configuration,  final site selection and foreground removal simulations. }

%%%%%%%%%%%%%%%%%%%%%%%%%%%%%%%%%%%%%%%%%%%%%%%%%%

\section{Introduction - HI intensity mapping and BINGO concept}

We are now firmly in the "era of precision cosmology", with a range of cosmological probes available that allow accurate measurements of cosmological parameters. However, with such precise measurements it is critical that we are not limited by systematic errors. One of the most important tools is the Baryon Acoustic Oscillations (BAOs) that provide a precise standard ruler (set by the sound horizon at the time of recombination) that we can use at much later times ($z<3$) to measure the expansion history of the Universe. Large optical galaxy surveys have now measured the BAO feature to $\approx 1\,\%$ precision \cite{Anderson2014}. The Square Kilometre Array (SKA), to be built in the next 10--15 years, will allow a similar analysis to be made. In the mean time, a more efficient and cheap method is available: HI intensity mapping \cite{Peterson2006}. The idea is to make a deep HI spectral survey at relatively low angular resolution ($\sim 1^{\circ}$) to map the {\it fluctuations} in HI brightness from many galaxies within the beam. 

BINGO\,\footnote{The Baryon acoustic oscillations In Neutral Gas Observations (BINGO) experiment is a collaboration between the University of Manchester and University College London in the UK, ETH in Zurich, Switzerland, University of Sao Paolo and INPE in Brazil, University of Montevideo in Uruguay and KACST in Saudi Arabia} is a novel and cost-effective way to map the large-scale redshifted hydrogen using the intensity mapping technique. BINGO will map the HI line over the redshift range $z=0.12$--0.48 (observing frequencies 960--1260\,MHz), with a mean redshift $z\approx 0.3$. This will allow a measurement of the acoustic scale to a few \%, thus providing an independent measurement of BAOs at radio wavelengths and constraints on cosmological parameters, such as the equation-of-state of dark energy. The guiding principle with BINGO is {\it simplicity}. We are using a single-dish total power design, with no moving parts. The dish(es) will be wire mesh and static, utilising the Earth's rotation to map a $\sim 15^{\circ}$ strip of the sky using a focal plane horn array. We will use a standard correlation receiver to reduce $1/f$ noise in the receivers, using the South Celestial Pole (SCP) as a reference to provide almost perfect matching of the sky and reference inputs. The detectors will use room-temperature Low Noise Amplifiers (LNAs) that should allow us to achieve a total system temperature $T_{\rm sys}=50$\,K. Much of the receiver will be in digital hardware, which is reasonably cost effective at these frequencies. The total cost is expected to be less than US\$4M and if funded (we are currently awaiting decisions on a number of funding proposals), we expect to be operational within 18 months. 

%%%%%%%%%%%%%%%%%%%%%%%%%%%%%%%%%%%%%%%%%%%%%%%%%%
\vspace{-1mm}
\section{BINGO update: Two-mirror design and site selection}

\subsection{Two-mirror design}
\label{sec:optics}

The original BINGO concept was to use a single 40\,m dish at the bottom of a large ($\approx 90$\,m) cliff edge \cite{Battye2013}. The horn array would be supported at the top of the cliff on a boom. This was believed to the most economical option while still meeting the requirements for beam performance and focal plane area to support $\sim$50 horns. However, we found it difficult to find an appropriate cliff, away from cites and facing roughly N-S (\S\,\ref{sec:site}). We considered using a large tower to house the horn array, but this would be very expensive. This opened up the idea of using a two-mirror design, since the cost of a second mirror would be of order the same as a tower.

\begin{figure}[!h]
\centering
\includegraphics[width=0.55\linewidth]{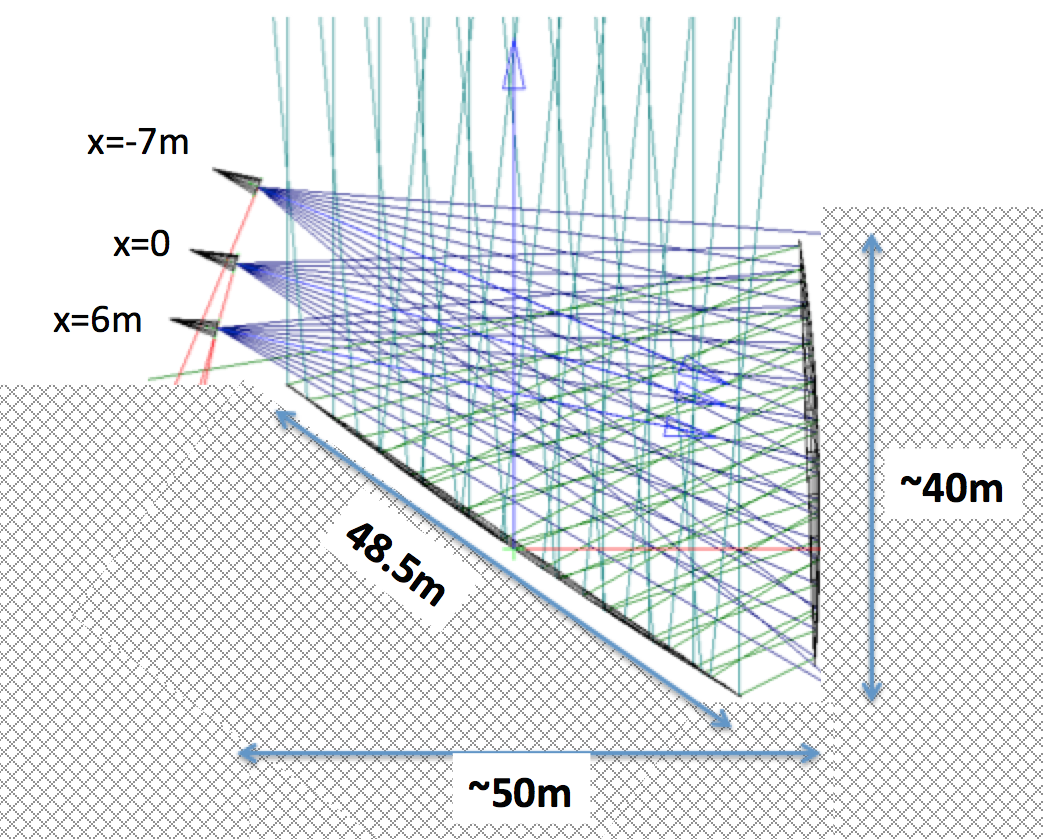}
\caption{Two-mirror crossed-Dragone design for BINGO. This design uses two $\approx 40$\,m dishes in a very compact configuration, which offers better optical performance than a single dish as well as reducing the size of the horns. It also makes more sites available (\S\,\ref{sec:site}). }
\label{fig:optics}
\end{figure}

The most obvious choice is the crossed-Dragone  (or Compact Range Antenna) configuration which has been extensively used by ultra-sensitive, wide field-of-view (FOV) CMB experiments \cite{Tran2010}. This design offers a very large FOV, with a flat focal plane that does not require refracting optics. The horn dimensions are also reduced to 1.7\,m diameter and 4.7\,m length, which is an important consideration given that the fabrication of 50 large metal horns is likely to be the greatest challenge and cost. Furthermore, the new optics offer exceptional beam performance, both in terms of sidelobe response and polarization purity; see Table~\ref{tab:optics}. In particular, the cross-polar response is exceptionally low even for horns at the edge of the array, which will mitigate polarized synchrotron radiation from leaking into the cosmological intensity signal.  Fig.~\ref{fig:optics} shows the new BINGO optics, highlighting the very compact nature of this optical configuration. Perhaps more importantly, it gives us many more options in terms of site selection (we only require a $\sim 40$\,m cliff and a slope; \S\,\ref{sec:site}). With some optimization, it may also allow us to cover a larger sky area and/or increase the number of horns to increase the overall sensitivity. 

\begin{table}[!h]
\caption{Representative performance figures for the optical performance of the two-mirror design as a function of offset from the optical axis.}
\label{tab:optics}
\begin{center}
\begin{tabular}{|l|c|c|c|}
\hline
\small Position &0\,m &6\,m &7\,m \\ \hline
Sky position &$0^{\circ}$   &$-5.2^{\circ}$   &$6.3^{\circ}$  \\
Optical gain      &51.68\,dB &51.45\,dB &51.22\,dB  \\ 
Peak cross-pol &$\textless -60$\,dB &$-43$\,dB  &$-41$\,dB \\
Beam ellipticity  &0.7\,\%  &4\,\%   &10\,\%  \\
\hline
\end{tabular}
\end{center}
\end{table}

\vspace{-1mm}
\subsection{Site selection}
\label{sec:site}

\begin{figure}[!h]
\centering
\includegraphics[width=0.45\linewidth]{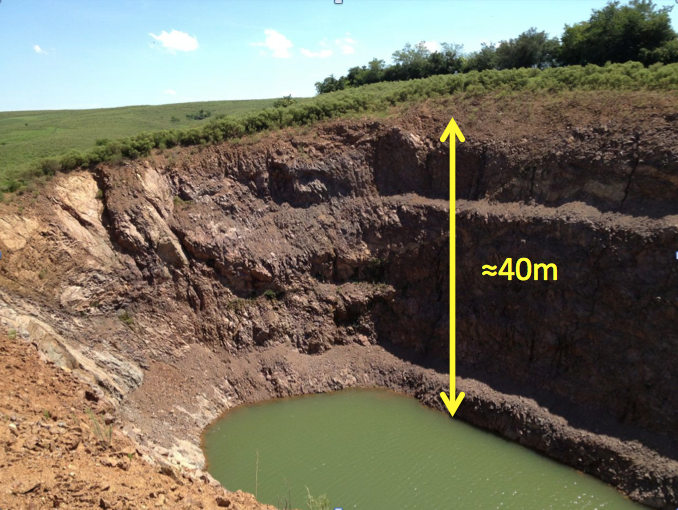}
\includegraphics[width=0.45\linewidth]{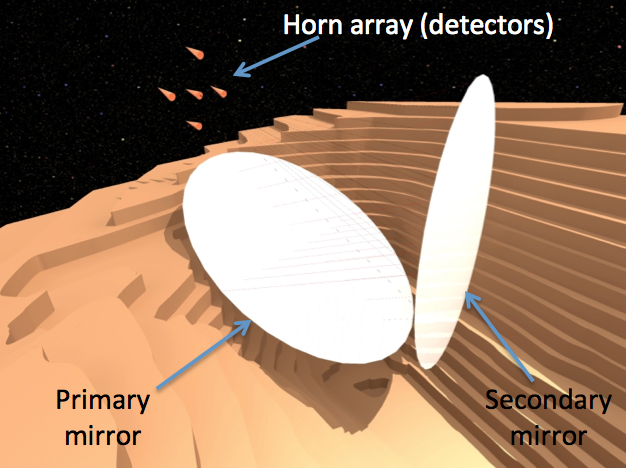}
\caption{{\it Left}: Photo of the Castrillon mine, which will be the site for the BINGO telescope. {\it Right}: 3-D model of the Castrillon site showing approximately how the two-mirror design will fit inside.}
\label{fig:site}
\end{figure}

The original concept of a $\sim 90$\,m cliff was challenging. With the new optical design (\S\,\ref{sec:optics}), we can afford to consider a number of sites. Due to the opportunities of funding in South America via Brazil/Uruguay, and the requirement to observe near the Celestial Equator (dec. $\approx -5^{\circ}$ is about ideal in terms of sky area and observing low-foreground regions) with reference horns pointing towards the SCP, Uruguay was our country of choice. We have recently surveyed several mines and valleys for Radio Frequency Interference (RFI), which have showed that the frequency band just above the GSM mobile phone band (above 960\,MHz) is indeed very clean. In the end, we found an unused mine in the North of Uruguay, near the town of Minas Corales. The mine is called "Castrillon" and is shown in Fig.~\ref{fig:site}. It features a large hole with a $\sim 45^{\circ}$ slope to the South, a $\sim 40$\,m slope to the North, and a flat area above the slope allowing the SCP to be observed. It is therefore well suited to the new BINGO configuration.\footnote{The site is not quite ideal in that it is not exactly N-S. Also, the water at the bottom of the mine needs pumping out but we believe this is a surmountable task!}

%%%%%%%%%%%%%%%%%%%%%%%%%%%%%%%%%%%%%%%%%%%%%%%%%%
\vspace{-1mm}
\section{Sensitivity and foreground removal}

\begin{figure}[!h]
\centering
\includegraphics[width=0.62\linewidth]{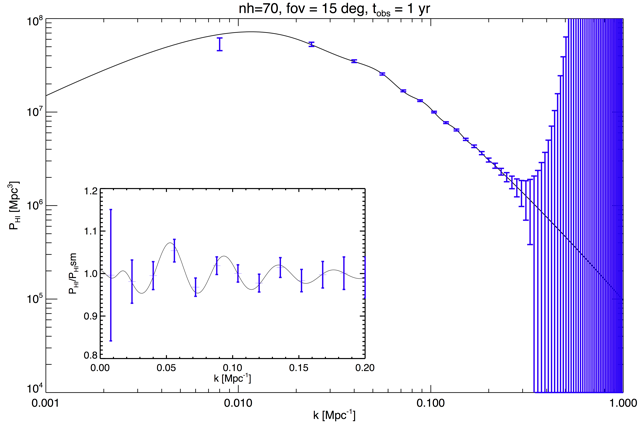}
\includegraphics[width=0.37\linewidth]{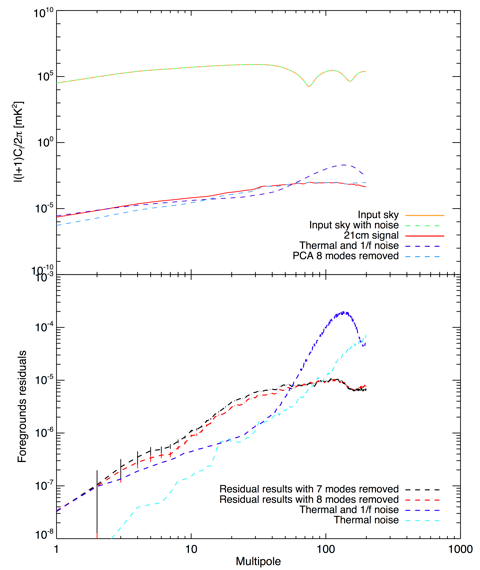}
\caption{{\it Left}: Projected power spectrum sensitivity for a full 1-year of BINGO observations, with 70 horns and $15^{\circ}$ FOV, providing a high S/N detection of BAOs. The panel insert highlights the BAO features after dividing out the smoothed spectrum.  {\it Right}: Application of PCA to remove foregrounds for simulated BINGO at 982\,MHz ({\it top panel}). The bulk of the foreground power has been removed close to the thermal noise level ({\it bottom panel}).}
\label{fig:sensitivity}
\end{figure}

If BINGO is deployed with at least 50 horns, with an average system temperature $T_{\rm sys} \approx 50$\,K, it will be able to map a large ($\approx 3000$ sq. deg.) region of the sky with a temperature sensitivity of $\approx 0.1$\,mK per 1\,MHz channel. This will allow a detection of BAOs at $\textgreater 5\sigma$. Note that the new optical design allows a larger FOV ($\sim 15^{\circ}$) than originally discussed \cite{Battye2013}, and potentially more horns, which would reduce the uncertainties, particularly on large angular scales. Fig.~\ref{fig:sensitivity} ({\it left}) shows the predicted sensitivity to the BAOs for a full 1-year observation (this would likely take $\sim 2$ years in practice) for 70 horns and $15^{\circ}$ FOV. This provides a measurement of the acoustic scale to about $2\,\%$, which is about the level of the current state-of-the-art large optical surveys. 

One of the most important considerations for HI intensity mapping experiments is foreground removal. The Galactic synchrotron radiation and emission from extragalactic sources is expected to be several orders of magnitude ($\sim 70$\,mK rms at 1\,GHz) above the cosmological signal ($\sim 0.1$\,mK rms). Although this presents a major challenge, we know that the foreground spectrum will be exceptionally smooth - to first order it is a power-law with some curvature. If the data can be calibrated sufficiently accurately, this will allow the foregrounds to be removed from the data since the cosmological HI signal is uncorrelated between frequency channels. 

A number of foreground removal techniques are available. A simple blind method is Principal Component Analysis (PCA), which decomposes the covariance matrix of the data into eigenmodes, of which only a small number contain the foreground (smooth) component. Fig.~\ref{fig:sensitivity} ({\it right}) shows an example power spectrum before and after applying PCA on simulated BINGO data, including thermal and $1/f$ noise. Virtually all of the foreground power can be removed to close to the thermal noise level by subtracting 8 modes; the foreground residuals are below the thermal noise level for $\ell \textgreater 100$. More detailed simulations of $1/f$ noise and systematics will be needed to quantify the level of residual foreground contamination.

%%%%%%%%%%%%%%%%%%%%%%%%%%%%%%%%%%%%%%%%%%%%%%%%%%
\vspace{-1mm}
\section{Conclusion and outlook}

BINGO is a novel and cost-effective way to map redshifted HI at $z=0.12$--0.48 with the aim of detecting BAOs at radio wavelengths. As well as providing independent cosmological data, it will have different systematics compared to optical BAO surveys. We are ready to start construction when funding has been secured\,\footnote{We are currently awaiting the outcome of funding proposals.}. The deep spectroscopic survey will have a wide range of scientific applications from cosmology to our Galaxy. With an upgrade to the digital backend, we may also be able to detect a number of Fast Radio Bursts (FRBs) of which only $\approx 10$ have currently been detected so far and whose origin is unknown.

%%%%%%%%%%%%%%%%%%%%%%%%%%%%%%%%%%%%%%%%%%%%%%%%%%

\section*{Acknowledgments}
CD acknowledges support from an ERC Starting Grant (no.~307209). We thank the Orosur mining company (Uruguay) for providing accurate drawings of the Castrillon mine.

%%%%%%%%%%%%%%%%%%%%%%%%%%%%%%%%%%%%%%%%%%%%%%%%%%

\end{document}